# Validation of human benchmark models for Automated Driving System approval: How competent and careful are they really?

Pierluigi Olleja [1]*, Gustav Markkula [2], Jonas Bärgman [1]

[1] Division of Vehicle Safety, Chalmers University of Technology, Gothenburg, Sweden
[2] Institute for Transport Studies, University of Leeds, Leeds, United Kingdom
*pierluigi.olleja@chalmers.se

## Abstract

Advanced Driver Assistance Systems (ADAS) and Automated Driving Systems (ADS) are expected to improve comfort, productivity and, most importantly, safety for all road users. To ensure that the systems are safe, rules and regulations describing the systems' approval and validation procedures are in effect in Europe. The UNECE Regulation 157 (R157) is one of those. Annex 3 of R157 describes two driver models, representing the performance of a "competent and careful" driver, which can be used as benchmarks to determine whether, in certain situations, a crash would be preventable by a human driver. However, these models have not been validated against human behavior in real safety-critical events. Therefore, this study uses counterfactual simulation to assess the performance of the two models when applied to 38 safety-critical cut-in near-crashes from the SHRP2 naturalistic driving study. The results show that the two computational models performed rather differently from the human drivers: one model showed a generally delayed braking reaction compared to the human drivers, causing crashes in three of the original near-crashes. The other model demonstrated, in general, brake onsets substantially earlier than the human drivers, possibly being overly sensitive to lateral perturbations. That is, the first model does not seem to behave as the competent and careful driver it is supposed to represent, while the second seems to be overly careful. Overall, our results show that, if models are to be included in regulations, they need to be substantially improved. We argue that achieving this will require better validation across the scenario types that the models are intended to cover (e.g., cut-in conflicts), a process which should include applying the models counterfactually to near-crashes and validating them against several different safety related metrics.

## 1 Introduction

During the last few decades, Advanced Driver Assistance Systems (ADAS) have been introduced in many markets and have in some cases become mandatory (Regulation 2019/2144). Such systems have been shown to reduce crash and injury risks substantially (Cicchino, 2017, 2018; Cox et al., 2023; Fildes et al., 2015). Meanwhile, Automated Driving Systems (ADS)—with a higher level of automation than ADAS—are being developed, and companies are releasing their first versions on public roads (Golson, 2023; Kusano et al., 2023). At higher levels of automation, here defined as SAE Levels 3-5 (L3–L5) on a scale from zero to five, the vehicle takes over the driving task completely. This capability is limited to specific driving conditions in Levels 3 and 4, and the driver may be requested to take back control in Level 3 (SAE, 2021). The expected benefits of ADS over ADAS include increased comfort, productivity, and safety (Bjorvatn et al., 2021; Wimmer et al., 2023).

As ADS are expected to take over the complete driving task (at least under specific circumstances) without having the driver as an imminent fallback, it is particularly important to ensure their safe operation from the first time they are released (without trained vehicle operators that supervise the driving task; Schwall et al., 2020) in real traffic. Therefore, a range of methods is employed to ensure the safe operation of ADS before they can be released for public use. Not only does the ADS industry run their vehicles through rigorous testing (ISO, 2022, 2023), but also authorities (at least in many countries) require ADS vehicles to be approved before allowing them on the road. For ADS, UNECE Regulation 157 (UNECE, 2023) is critical: all ADS vehicles in Europe that have an Automated Lane Keeping System (ALKS) need to conform to this regulation before they are allowed on highways (Regulation 1143/2014). One of the regulation's components introduces two (independent) "competent and careful" reference drivers (Annex 3 in UNECE, 2023, p. 40) from which "guidance can be taken" to define whether cut-ins, cut-outs, and decelerating lead-vehicle scenarios are preventable or unpreventable. In a cut-in, a leading vehicle in an adjacent lane (hereafter called principal other vehicle; POV) encroaches into the lane occupied by the ego vehicle (i.e., the vehicle with the ADS or reference driver model) with a short headway. In a cut-out, a vehicle in front of the ego vehicle leaves the lane to reveal a slower (or immobile) vehicle in the ego vehicle's lane. The decelerating lead-vehicle situation is a common rear-end crash conflict in which the lead vehicle brakes unexpectedly. The UNECE assessment of ALKS includes these three scenarios.

The UNECE reference driver models are mathematical representations of human drivers. The assessment (to determine whether an ADS is safer than each of the two models) is operationalized through computer simulations. That is, the ADS and the reference driver model are both applied to a set of digital representations of different traffic situations. One of the UNECE pass/fail criteria is to only consider the ADS safe enough if it ensures a performance "at least to the level at which a competent and careful human driver could minimize the risks" (UNECE, 2023; p.16). That is, the reference models are used as benchmarks of reasonable human responses, and are hereafter referred to as human benchmarks.

The two UNECE models are a) the "competent and careful" driver model (CCDM; JAMA, 2022) and b) the Fuzzy Safety Model (FSM; Mattas et al., 2022). In the regulation they are called "Performance model 1" and "Performance model 2", respectively. As mentioned, the two models are intended to guide the definitions of driving situations as either preventable or unpreventable. The CCDM uses kinematic thresholds (e.g., Time Headway, Time-To-Collision (TTC), and the position of the ego vehicle and POV in their lanes) as part of the threat assessment. The FSM uses fuzzy logic (Mattas et al., 2020; Zadeh, 1965) to produce a response that is proportionate to the level of the threat, and to detect the risk of collision early. Both models can only react longitudinally (by braking).

As the UNECE models are part of a regulation, they affect several stakeholders—from the organizations that need to comply with the regulation to society as a whole. For the latter an important question is: Have the vehicles complying with the regulation been compared with a "competent and careful" driver? The answer is relevant for traffic safety.

To our knowledge, there has been only one study that assessed the UNECE models. Mattas et al. (2022) applied the UNECE models to naturalistic vehicle trajectory data from the highD drone dataset (Krajewski et al., 2018). We have found no publications on the application of the UNECE models to safety-critical events—either crashes or near-crashes. This means that today it is unknown how the UNECE models perform in safety-critical events: Do they perform as "competent and careful" human drivers? This study focuses on the UNECE models' performance in cut-in near-crashes.

The rationale for choosing to assess the UNECE models for cut-ins specifically is twofold: a) availability of data, and b) it is more difficult to develop driver behavior models for cut-ins than for traditional same-lane rear-end scenarios. As for a), this work was done within a project that has access to cut-in near-crashes from the SHRP2 naturalistic driving dataset (Hankey et al., 2016; p. 7). Further, SHRP2 includes a relatively large number of cut-ins, but few cut-outs. As for b), rear-end modeling has been going on for many years (see Markkula et al., 2012, for a review), while cut-in modeling is not as common. Less research has been done to understand and model how a competent and careful driver would predict and react to an imminent crash or conflict in a cut-in scenario than in a pure rear-end scenario. For example, a key challenge in modeling cut-ins is to algorithmically describe when the ego driver considers a lane change to have been initiated by the other driver (Jokhio et al., 2023; Kauffmann et al., 2018). Additionally, other aspects of the response process, such as when and how hard the driver brakes, have not been studied as extensively for cut-ins as they have been for rear-ends.

In sum, no other work assesses the validity of the UNECE reference driver models as "competent and careful" when applied to near-crashes. This work aims to address this research gap specifically for cut-in near-crashes, by assessing the models' safety performance when they are applied to the kinematics of real world cut-in near-crashes from the SHRP2 naturalistic driving study. Three metrics from the SHRP2 data (crash avoidance performance, the timing of braking reactions, and the other vehicle's lateral position when the human drivers brake) were compared with results from the CCDM and the FSM. We also reflect on reference driver modeling in general and its challenges.

## 2 Method

### 2.1 Data, scenario description, and pre-processing

This study used near-crash cut-in events from the Strategic Highway Research Program 2 (SHRP2) database. SHRP2 is a large Naturalistic Driving Study (NDS) that collected data from more than 3000 volunteers driving on public roads over a period of two years in six regions in the USA (Blatt et al., 2015). The vehicles were equipped with a front radar and several cameras, including a front-facing camera and a camera facing the driver. In addition to sensors such as accelerometers, angular rate sensors, and global positioning systems (GPS), some data from the vehicle electronic bus system (Controller Area Network; CAN) were also collected. In this study the front-facing camera, the front radar, and the ego vehicle speed from the CAN were used.

As part of the SHRP2 project, safety-critical events were identified through kinematic triggers (e.g., driver deceleration; see Hankey et al., 2016, for trigger details), extracted, reviewed by expert annotators, and categorized by level of criticality (Hankey et al., 2016). The safety-critical cut-ins in this study can be described as follows: an ego vehicle travels straight on a road with more than one lane in its direction of travel. A POV travels in the same direction as the ego vehicle in a lane adjacent to the one occupied by the ego vehicle, but at a lower speed. The POV then performs a rather abrupt lane change ahead of the ego vehicle, requiring a sudden evasive maneuver by the driver of the ego vehicle (for details see Fig. 1A and the SHRP2 definitions of near crash and POV lane change in front of the ego vehicle in Hankey et al., 2016, p. 7; VTTI, 2015, p. 30).

The current analysis is a follow-up on a previous study (Chau & Liu, 2021), which tried to computationally model the ego drivers' responses to safety-critical cut-ins. We will therefore explain their approach to selecting which events to analyze, which proceeded as follows: 685 events were extracted from the SHRP2 database, using the filter function in the SHRP2 insight database (VTTI, 2024). The filtering criteria that resulted in 685 events are described in the Appendix, in Table 1. Detailed time-series data for these events were obtained from VTTI through a data license agreement.

Only three of the events were crashes (<0.5%). Further, the vast majority of the events (632) were categorized in the SHRP2 database as "Conflict with vehicle in adjacent lane" near-crashes, and the remaining 50 near-crashes as "Conflict with merging vehicle". From these 685 events, 209 near-crashes were selected for further review. The limit of 209 was set by the resources available for review in the original study, and the events chosen were simply the first 209 events provided from the filtering, excluding events where it was obvious the data were not complete (e.g., where video was missing). The 209 events where then further reduced to 51 (Chau & Liu, 2021). Specifically, cut-ins at very low absolute speeds (e.g., while in a traffic jam), cut-ins in which the ego driver reacted to the threat by steering instead of braking (the UNECE models are only braking), cut-ins that were not considered unsafe (e.g., large longitudinal time gaps between the ego vehicle and the POV, and similar speeds), cut-ins where the POV was not visible for long enough in the front video to allow for proper analysis, cut-ins with the vehicles traveling on a road with more than a slight curvature, and events with data quality issues (in addition to those obvious at first review), were excluded. Note, however, that as this was done in a previous study, we do not have full documentation of the exact number of excluded events for these reasons for exclusion. Chau and Liu (2021) also excluded events where distraction was annotated in SHRP2 as a contributing factor to the safety-critical event (e.g., near-crash or crash), in order to focus on eyes-on-threat response modeling. This exclusion is also appropriate for the scope of the current work, as the UNECE models are supposed to represent a competent and careful driver, which can be taken to mean drivers that are not distracted (UNECE, 2023).

For the current study, another 11 events were excluded during the annotation process, either because some feature of the POV was occluded or because the video quality was too poor for the more detailed reconstructions carried out here. Additionally, in two of the original events it was not possible to identify the beginning of the evasive maneuver (e.g., in one event the threshold level of deceleration was not reached, and in the other the ego vehicle was already decelerating when the annotated event began), therefore these events were also excluded from all analyses. In total 38 events were finally used in the analysis.

For each of the 38 events, the kinematics of the two involved road users (ego vehicle and POV) were reconstructed using data from the ego vehicle. Specifically, the cut-in events were annotated to reconstruct the POV trajectory, using an annotation tool initially developed by Shams El Din (2020; see Fig. 1B). The post-processing of the annotation tool output was further developed in this work. The front-facing camera and the radar were used to detect the relative lateral and longitudinal distances between the vehicles and the heading angle of the POV in relation to the lane. The annotation tool enabled the user to manually place a bounding box around the POV in order to determine the position and angle of the POV from the front-facing camera view only. The reason for not using the radar data was that the SHRP2 radar data are of relatively low quality: there are gaps in the data, multiple vehicles are sometimes tracked at once and sometimes "mixed up", and data are missing altogether for some events. However, the radar data can be used to improve the camera-based distance estimation: for each event with radar data available, the camera-estimated distance to the POV was compared with the radar signal when the ego vehicle and POV were at their closest. (The radar signal is most reliable when the ego vehicle and the POV are closest, both laterally—due to limited horizontal field of view—and longitudinally—due to quality issues.) If the user of the annotation tool identified a difference between the radar distance and the camera-estimated distance, this offset was then added or subtracted to the camera-estimated distances throughout the event.

The tool included an interpolation feature between video frames, providing the annotator with a POV bounding box overlay to verify that annotations and interpolations were correct (and if they were not, the annotator could add to or correct the bounding box annotation). In addition to the annotations for distance and angle estimation, the left and right lane marking of the ego vehicle's lane were manually annotated, to enable the estimation of the lateral offset of the ego vehicle from the center of its lane, and the distance from the POV to the lane marking. Particular attention was given to the moment when the POV touches the lane marking during the lane change maneuver: visual confirmation from the original video was used to ensure the timing of the lane touch in the original and the reconstructed events was the same.

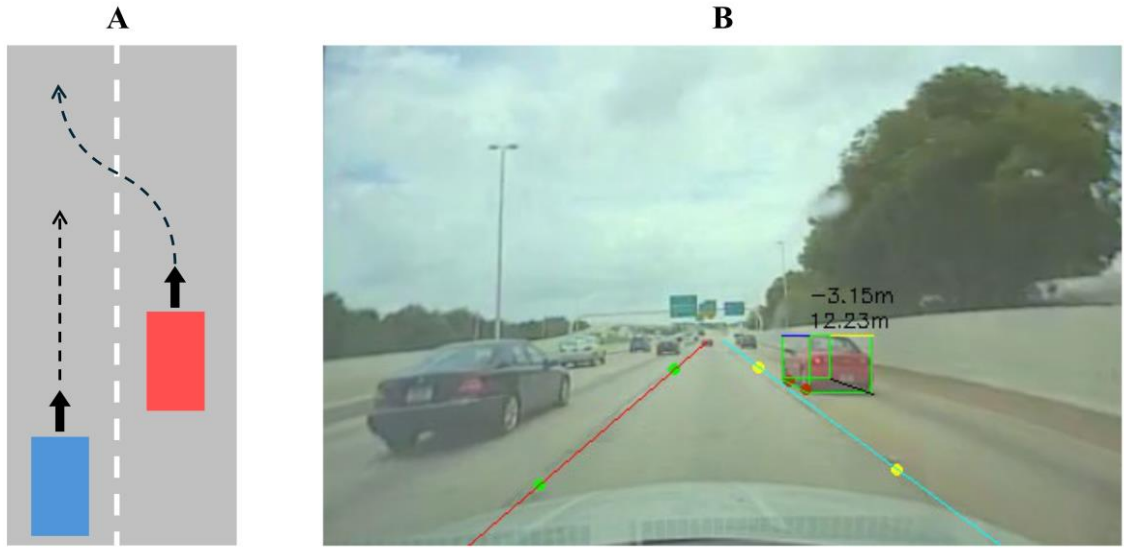

*Fig. 1 – Top-view schematic of a cut-in event (A). Annotation tool (B): the picture shows the bounding box of the POV, its contact point to the road, the lane markings, and the lateral and longitudinal relative distances calculated by the annotation tool, based on the annotations. Included with permission from VTTI.*

## 2.2 The UNECE models

The characteristics of the two UNECE models are described briefly in this section. Details about the models can be found in the UNECE regulation (UNECE, 2023).

*2.2.1 The competent and careful driver model (CCDM)*

Annex 3 in the regulation describes the CCDM, including model parameter values. Note, however, that the regulation does not provide the sources or rationales for the selection of those specific values. However, JAMA (2022), provides some background to the CCDM and the parameter value choices. The following is a short overview of the model, including the parameter values.

Although the UNECE model covers three types of situations, this work focuses exclusively on the cut-in scenario. For this scenario, UNECE defines a "wandering zone", the area within which the POV can move laterally without the ego vehicle taking any evasive actions (basically delineating normal lane-keeping by the POV). This area is defined as 0.375 m to the left and right of the center of the POV lane. In the UNECE model, if the POV leaves the wandering zone (that is, if it moves more than 0.375 m away from the center of the lane), the driver (model) of the ego vehicle in the adjacent lane starts perceiving the cut-in, followed by a reaction time, which is the sum of a risk perception time (0.4 s) and a braking delay (0.75 s). Once the reaction time has passed, the ego vehicle starts braking, but only when/if the longitudinal time to collision (TTC) is less than 2 s. In the UNECE model, the TTC is defined purely longitudinally (note that they did not describe the details of the TTC calculations). If the POV is ahead of the ego vehicle, TTC is positive, while if the POV and the ego vehicle are overlapping longitudinally (in adjacent lanes), TTC is negative. The model uses a constant deceleration of 7.6 m/s$^2$, reached 0.6 s after the braking onset (i.e., a jerk of 12.65 m/s$^3$).

*2.2.2  The Fuzzy Safety Model (FSM)*

The FSM describes a driver that "can anticipate the risk of a collision and apply proportionate braking" (UNECE, 2023; p. 56). The decision and action process of the FSM is divided into three main steps: a lateral safety check, a longitudinal safety check, and a reaction. The lateral and longitudinal checks evaluate the risk of encroachment onto the future path of the ego vehicle by the POV. If these checks indicate the possibility of an unsafe situation developing, the model computes two metrics to assess the need and the intensity of a reaction by braking. The FSM uses fuzzy logic as opposed to clearly defined static thresholds for threat assessment. Mattas et al. (2022) state that this logic enables a response that is proportional to the threat—for example, "calm deceleration". The first metric is the Proactive Fuzzy Surrogate metric (PFS), intended "to identify situations where a vehicle is not driving in a defensive, careful manner", and the second metric is the Critical Fuzzy Safety metric (CFS), used "to identify situations where an accident is imminent and action should be taken so the accident is avoided" (Mattas et al., 2019; p. 2). The PFS and CFS can have values between 0 and 1. When CFS = 0, the intensity of the braking reaction increases linearly with the value of the PFS, from 0 m/s$^2$ up to the comfortable braking deceleration of 4 m/s$^2$. If CFS > 0, the intensity of the deceleration can reach up to 6 m/s$^2$. The jerk is 12.65 m/s$^3$, as for the CCDM. For further details about the FSM (and, specifically, the PFS and CFS), see work by Mattas et al. (2020), Mattas et al. (2022) and UNECE (2023).

## 2.3  The Counterfactual simulations

The virtual simulation toolchain used for the simulations in this study was based on the esmini simulation tool[1]. The UNECE model implementations were based on open-source versions of these models[2], compatible with esmini.

The driver models were applied to all 38 SHRP2 cut-in events after any evasive maneuver by the real driver in the original event was removed (Bärgman et al., 2017). The onset of the human driver's evasive maneuver for each SHRP2 event was defined as the first instant in which the ego vehicle deceleration reached -0.2 m/s$^2$ in the unmodified (original) pre-crash kinematics. From this point on, the ego vehicle speed was kept constant (see Fig. 2). The driver models were then independently applied to these modified events.

For each simulation (i.e., when each individual driver model was applied to an individual modified event) the POV followed its original trajectory, and the ego vehicle followed the modified trajectory until the driver model "requested" braking. At this point the model took over control of the ego vehicle. The POV always completed its original trajectory, ignoring any actions of the ego vehicle.

To get a worst-case reference, all modified events were also simulated without any active UNECE models (i.e., without any ego driver reaction).

---

[1] https://github.com/esmini/esmini

[2] https://github.com/esmini/esmini/tree/master/EnvironmentSimulator/Modules/Controllers

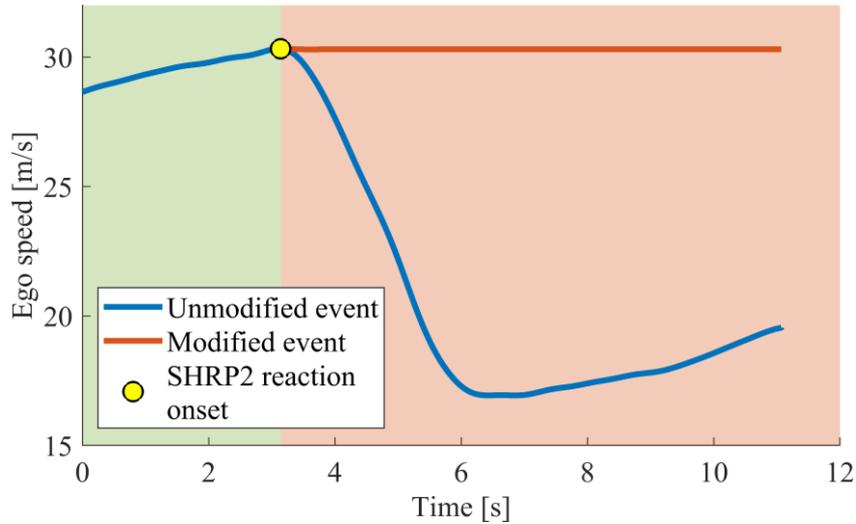

*Fig. 2 – An illustration of the speed profiles of an unmodified cut-in event and its corresponding modified event—the latter being used in the counterfactual simulations. If an applied UNECE model initiates braking to the left of the SHRP2 reaction onset (green area), the model responds earlier than the human driver did, and if it initiates braking to the right (pink area), it responds later than the human driver did.*

## 2.4 Analysis

Three metrics were used to compare brake initiation by a human driver with the CCDM and FSM models: response time difference ($t_{diff}$), crash avoidance (yes/no), and the POV lateral distance at brake onset (LDBO, in absolute values and as a difference between model and human response). These three metrics are described in turn.

The variable $t_{diff}$ represents the models' reaction time minus the human drivers' reaction time. A negative value of $t_{diff}$ means that the model reacted earlier than the human driver in the original event, while a positive value means that the driver model reacted later. This metric compares the responsiveness of the models and the human driver.

Crash avoidance was determined by simply checking if the UNECE models generated any crashes, and, if so, how many. All the original events were near-crashes, so as mentioned earlier, the expectation was that neither of the models would generate any crashes.

The third and final metric used in the comparison was the LDBO. This metric is an indication of whether, and by how much, the POV is intruding into the ego vehicle's lane when the ego driver starts braking. Analysis included both the absolute value of the POV's distance from the lane marking and the differences between this distance for the human driver and each of the models (to assess the similarity between the human and the models). As seen in Fig. 3, the distance to the lane marking was calculated from the corner of the POV that was closest to the ego vehicle.

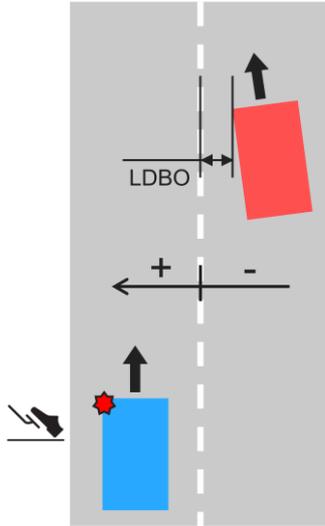

*Fig. 3 – An illustration of the lateral distance at brake onset (LDBO) metric. If the ego driver reacts when the POV is still in its initial lane, LDBO is negative. LDBO increases as the ego driver delays its reaction.*

## 3  Results

### 3.1  Timing of the evasive maneuver and crash avoidance

The two driver models were individually applied to the 38 safety-critical near-crash cut-ins from SHRP2. The CCDM analyses include only 34 events because the model did not react in four events: in three events the cut-in was detected but the braking maneuver was not triggered because when the reaction time had "passed", the kinematics of the event were such that TTC was greater than the 2 s threshold set by the model, and increased further until the end of the event (the POV accelerated during the cut-in maneuver); in one event braking was not triggered because the CCDM detected the cut-in when the ego vehicle had already started overtaking the POV (the kinematics of the event had changed due to the process described in Section 2.3). None of the four events in which the CCDM did not react resulted in a crash. The FSM reacted in all the available events; hence the FSM analyses include all 38.

Results (Fig. 4A) show that the CCDM response was delayed compared to the human drivers in the majority of the events (27 out of 34: 79%): the median $t_{diff}$ was 0.5 s. Although there were originally no collisions in any of these 34 SHRP2 events, the CCDM crashed in three of them (9%). To understand the root cause of the late response of the CCDM, the timing of its evasive maneuver onset was further analyzed. As described previously, the CCDM begins its "perception time" (followed by the reaction time) only after the POV exits the wandering zone, and does not perform the evasive maneuver until the TTC is lower than 2 s. The time that the POV takes to reach the ego vehicle's lane and enter its path depends on the lateral speed of the POV.

In addition to analyzing the performance of the models, we also investigated what would happen if the driver in the ego vehicle would not react at all. These worst-case simulations resulted in 24 crashes out of the 38 events. Note that this means that there were 14 events with no physical contact between the ego vehicle and the POV, even without any evasive maneuver by the ego vehicle. Consequently, the CCDM crashed in 13% (three out of 24) of the worst-case crash events; the remaining 14 do not contribute to the assessment of the models' crash avoidance performance.

The time difference $t_{diff}$ of the FSM brake initiation is shown in Fig. 4B. The results show that the FSM reacted earlier than the human driver in the majority of cases (28 out of 38: 74%); the median $t_{diff}$ for the FSM is -0.7 s. The FSM did not generate any crashes. Note that positive $t_{diff}$ values for this model are generally shorter than the negative values. The reason for this is that in the events where a lack of reaction causes a collision (the 24 worst-case events described above), positive values are constrained by the collision time, while for negative values there is theoretically no such constraint (although practically they are constrained by the duration of the annotation).

Two-sided Wilcoxon signed-rank tests show that the brake onset times have a median significantly different from zero (i.e., the human drivers) with an α of 0.01 for both models (CCDM: W = 462, p = 0.0012; FSM: W = 116, p = 2.2e-04).

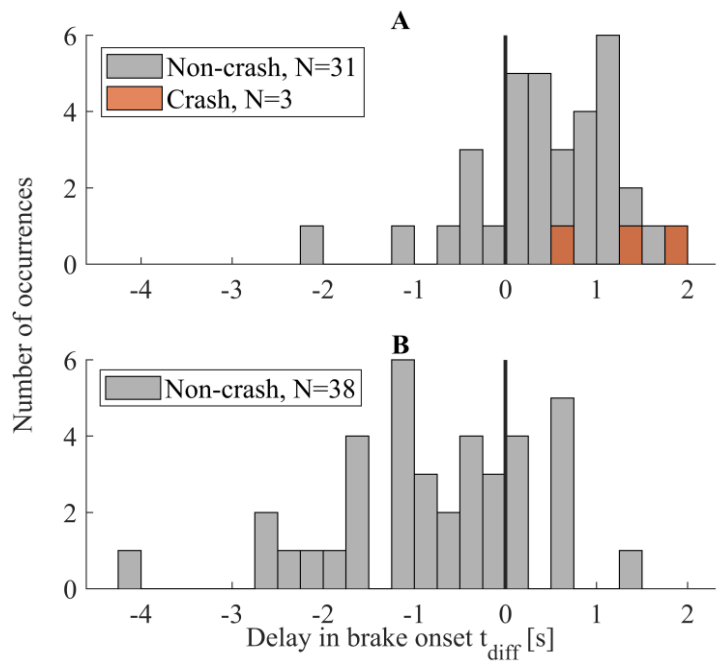

*Fig. 4 – The difference ($t_{diff}$) between the brake initiation of the human drivers in in the original events and the CCDM (A) and FSM (B). The value is positive if the model reacted later than the real driver, and negative if it reacted earlier. The bin width is 0.25 s. The red (dark) in the upper panel shows the time from the human driver's brake onset until the CCDM initiated the avoidance maneuver for the three events that resulted in crashes (due to late initiated avoidance).*

## 3.2 The distance to lane marking at brake onset

The results of the analysis of the LDBO metric are shown in Fig. 5. Fig. 5A shows that 63% of the human drivers reacted while the POV was still entirely in its initial lane. The median distance was -0.23 m. The CCDM (Fig. 5B) reacted when the POV was still in its initial lane in 35% of the events, less often than the human driver. The median of the LDBO for the CCDM was 0.24 m. The FSM (Fig. 5D) reacted before the POV started crossing the lane in 71% of the events, with a median of -0.30 m.

Further, the CCDM started braking later than the human driver—when the POV was laterally closer to the ego vehicle—in 74% of the events. (Note that Fig. 5C only includes the 34 events in which the CCDM reacted.) Unlike the CCDM, the FSM's braked when the POV was laterally closer to the ego vehicle (Fig. 5E), compared to the human driver's reaction, in only 26% of the events.

Two-sided Wilcoxon signed-rank tests between the LDBO for the human drivers and the CCDM (W = 92, p = 7.6e-04) and FSM (W = 567, p = 0.0044), respectively, showed significant differences at an α of 0.01.

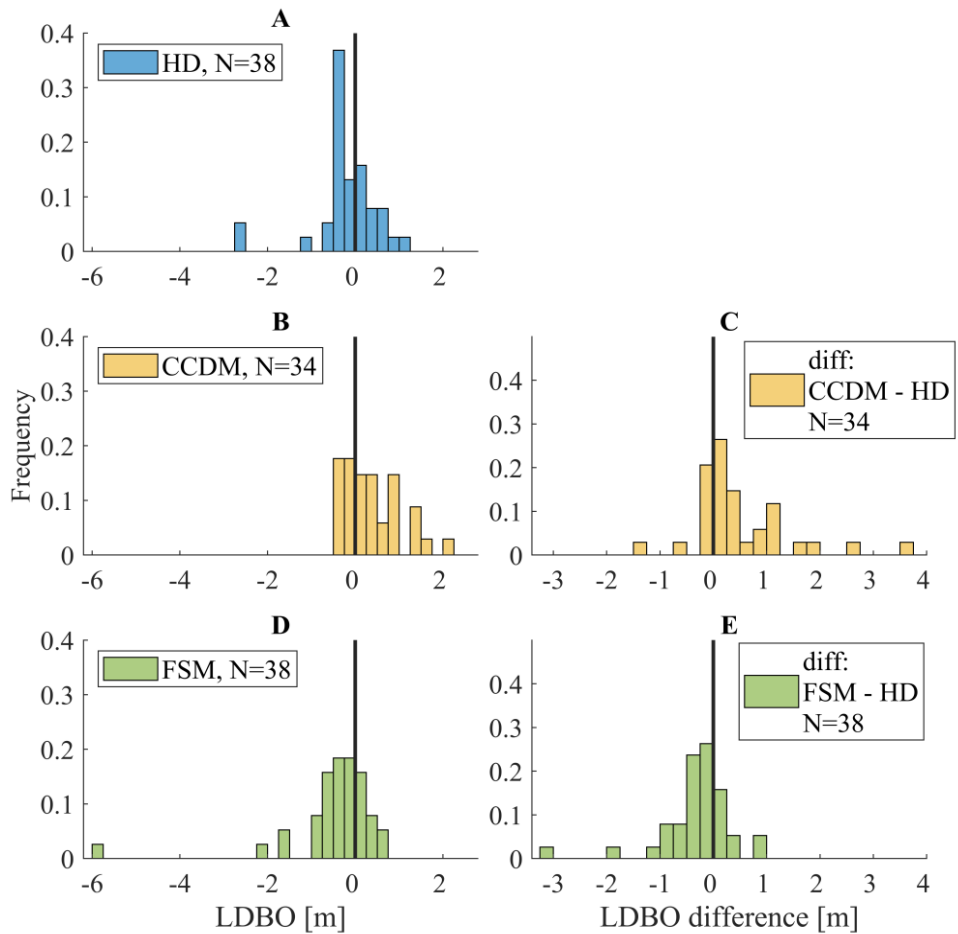

*Fig. 5 – Analysis of the lateral distance at brake onset (LDBO) metric. The bin width is 0.25 m. Panels A, B, and D show the LDBO, where a negative value indicates that the ego driver reacted when the POV was still fully in its lane. Panels C and E show the differences between the human driver (HD) and the models; a positive value means that the model reacted later (when the POV had moved further into/towards the ego vehicle's lane) than the human driver.*

## 4 Discussion

This study used counterfactual simulations to compare the two computational human-benchmark driver models described in UNECE (2023) with real human responses in cut-in near-crashes. The main take-aways from this work are that, in near-crash situations, the "competent and careful" driver model (CCDM) seems to be far from a "competent and careful" driver, braking substantially later than the human drivers, and that the FSM initiates braking substantially earlier than human drivers. While the human drivers did not crash in any of these near-crash events, the CCDM crashed in three of the events, or 13% of the events where a crash would have occurred if the original human driver had done nothing.

## 4.1 What are we comparing?

The CCDM and FSM are used in the UNECE Regulation 157 to assess whether a collision is preventable by human drivers in safety-critical events (UNECE, 2023); they are not intended to be used as controllers for uneventful driving scenarios (e.g., driving scenarios that don't involve safety-critical events; Mattas et al., 2022). This means that the type of data used is an important aspect of model validation. Mattas et al. (2022), that assessed the UNECE models, used only everyday driving NDS data. While such data provide large amounts of data for testing driver models, they only assess one aspect—how the model reacts to everyday driving. On the other hand, safety-critical event databases with measured crashes and near-crashes (such as SHRP2) can likely provide accurate (enough) representations of real-life situations for assessing the models. Olleja et al. (2022) demonstrated the limitations of normal driving data compared to near-crash data (for rear-end crashes) for capturing criticality. The situation should be comparable for cut-in scenarios. With this as a background, we argue that the set of events assessed here is well within the scope of the UNECE regulation: they are safety-critical, real-world events that real-world ALKS would have to deal with.

Before discussing the results of this study, it is important to introduce the biases of the study that could possibly have an effect on the results. The possible biases stem primarily from the original SHRP2 data collection and from our event-selection process. One example of bias in the SHRP2 data is that there were more younger and older drivers than in the overall US population of licensed drivers: the full SHRP2 dataset has 51% younger than 35 and 38% younger than 25, while for US licensed drivers the same numbers are 32% and 13%, respectively (Flannagan et al., 2019). However, the age bias of the full SHRP2 is smaller than in the dataset used for this study—at least for the proportion of drivers below 35: 66% of the SHRP2 drivers in this study were younger than 35, and 34% younger than 25.

Arguably, this bias towards younger drivers should mean that the drivers in this sample are less experienced, and therefore detect hazards "less quickly and efficiently" compared to more experienced drivers (Deery, 1999; p. 229). This likely means that SHRP2 drivers are not in general "competent and careful"—at least not with respect to the impact of age. It is also worth noting that SHRP2 consists of data from the US, which has a higher crash rate and different infrastructure than Europe overall (WHO, 2023). However, both crash rates and infrastructure vary across Europe (Eurostat, 2023), and the EC regulation that includes the UNECE models (UNECE, 2023) is valid across all those differences.

Another factor that impacts the representativeness of the SHRP2 data is the triggers used to identify the near-crashes. That is, an unknown is here how the initial triggering affected the dataset: as the 682 near-crash events were identified by kinematic triggers (as part of the SHRP2 project), if the ego driver did not react with harsh braking or swerving the event was not included in the original dataset (with some exceptions, given that some events could be flagged by the SHRP2 drivers themselves, or identified by data analysts while reviewing the video footage 'by chance'; Hankey et al., 2016). The reason for the lack of harsh braking may be that the ego driver missed the event altogether (e.g., due to distraction), but it went well even without any action at all by the driver (recall that 37% of the events in our simulations did not crash even without any action by the ego driver). The exclusion of safety-critical events completely missed by the driver would mean a bias in our dataset towards safer drivers. It may, on the other hand, be that the ego drivers noticed the situation early and started to adapt the vehicle speed early, thus not having to act with harsh braking or swerving. Exclusion of such events would then mean a bias towards less safe drivers. As we do not know anything about "non-triggered events", we cannot establish if the overall bias is towards safer or less safe drivers, but we believe the bias is towards less safe drivers (i.e., it is more likely that drivers avoid safety-critical situations without harsh braking than that they miss them completely and avoid them anyway).

The potential biases introduced so far have been about the original SHRP2 dataset. Our events selection process is, however, another potential source of bias that deserves attention. First, given that the following conditions likely apply consistently across all driver types, we make the assumption that the following exclusions did not introduce any bias: a) events with missing or low-quality data, b) events involving very low absolute speeds for both vehicles, c) events occurring on roads with larger curvature, and d) the initial random selection of 209 out of the 685 events. However, different aspects of the event selection process may have biased our sample both towards inclusion of safer and less safe drivers. As a first example, the fact that we exclude crashes should bias the data towards safer drivers. However, as crashes accounted for less than 0.5% of all events considered for inclusion (i.e., 99.5% of the original 685 cut-in events were near-crashes), the effects of this bias are probably minimal. As a second example, we also excluded events that did not seem very critical according to the video—even if they were classified as near-crashes in the SHRP2 annotations. Although as previously mentioned we do not have the exact number of events that were excluded for this reason, the number was far greater than the number of excluded crashes. Excluding these lowest-severity events in the dataset is likely to bias the data towards more risky drivers, but to what extent is unknown. As a third example, we excluded events where the SHRP2 annotation listed distraction as a contributing factor, a decision which could introduce a bias toward safer drivers. However, it is important to note that distraction was classified as contributing to only 6.3% of the full set of 685 cut-in events. For comparison, Balint et al. (2020) found that visually distracting secondary tasks in SHRP2 baseline driving segments had a prevalence of at least twice that percentage. Although secondary task prevalence and distraction may not be directly comparable, this comparison can be seen as an indication that the drivers in the SHRP2 cut-in near-crashes seemed to exhibit a relatively low level of distraction. If drivers are paying attention, then the POV is probably the primary contributor to conflicts.

In summary, we cannot with certainty determine whether the SHRP2's and our selection processes biased the dataset towards safer or less safe drivers. However, we believe that the dataset is closer to describing the average population, or even drivers driving less safe than average drivers, than to describing "competent and careful" drivers. If that is so, then for the majority of the assessed events, if the CCDM and FSM were truly models of "competent and careful" drivers, the models should respond earlier, or maybe even substantially earlier (but then as a clear response to the cut-in), than the human drivers in our dataset. In the next section, we discuss these timing issues further.

## 4.2 Evasive maneuver response timing

The main finding of this work is that the CCDM and the FSM handle cut-in scenarios quite differently from each other—and differently from the human drivers in the SHRP2 near-crashes. Specifically, the CCDM responded later than the human drivers in 79% of the events (a median of 0.5 s later). One likely explanation for the late response is that the CCDM relies only on lateral position (the boundary of the wandering zone) to detect the POV's lane change initiation. In engineering applications, a lane change is often defined using thresholds for the vehicle's lateral speed and position within the lane (Jokhio et al., 2023; Mullakkal-Babu et al., 2020; Wang et al., 2019). However, it can be difficult to find the right balance between correctly assessing the urgency of an imminent lane change and recognizing the natural lateral movements of a vehicle within the lane (satisficing in lane-keeping; Summala, 2007). A second and complementary explanation is that the perception and reaction times of the CCDM remain fixed, regardless of how swiftly the POV lane change is executed. Supporting these two explanations, a detailed case-by-case analysis showed that the three crashes in the CCDM simulations occurred when the POV had a relatively high lateral speed. That is, by only triggering on lateral position and with a fixed response delay, the CCDM is not able to avoid these three crashes, even if the humans in the original events avoided crashing.

Unlike the CCDM, the FSM uses the predicted motions (lateral and longitudinal) of the POV to predict whether the ego vehicle and the POV are on a collision path and triggers braking accordingly. Our results (Fig. 4) show that, for a majority of cases, the FSM responded earlier than the human drivers, with a much larger range of response times. Analysis of the events shows that this model tends to be more sensitive than the SHRP2 human drivers at predicting cut-in maneuvers, and consequently initiates braking earlier. In five events, the FSM reacted more than 2 s earlier than the human drivers. These results are in line with those in the study by Mattas et al. (2022). In that study the CCDM and the FSM were applied to normal, everyday highway driving events (using the highD dataset; Krajewski et al., 2018). They concluded that, in many events, the FSM triggered even when the vehicles in the adjacent lanes never crossed the lane marking. One can argue that a "competent" driver would not initiate braking for every small lateral motion of a vehicle in the adjacent lane, just because in some possible future (possibly several seconds away) there may be a collision.

In addition to analyzing the human and model responses in the time domain, we also examined the POV's lateral distance to the lane at the onset of braking (LDBO). The CCDM tended to initiate braking at a lower relative lateral distance in the majority of cases (Fig. 5C). Conversely, the LDBO for the FSM was generally similar to that of human drivers. These results suggest that, for the FSM, small lateral movements of the POV early in the event, well before the lane change initiation, triggered an intervention. Thus the FSM is more likely to trigger during uneventful driving than the CCDM is. Another aspect of the FSM worth noting again is that it triggers the braking even at large lateral distances between the ego and the POV, as long as the other conditions, such as path predictions, indicate a potential future collision. Actually, in one event, the FSM triggered braking when the POV was in the lane beyond the adjacent lane.

In summary, because the CCDM lacks a lateral urgency component, we argue that it cannot be considered a "competent and careful" driver in near-crash situations. For FSM, on the other hand, there are indications that it is overly sensitive to lateral perturbation by the POV. Further, the lack of an urgency component can make the FSM trigger when there is a relatively large separation between the ego and the POV in both time and space—arguably larger than in a reasonable human response, even from a competent and careful driver.

### 4.3 Human benchmark models for ADS assessment

In this study, the CCDM was shown to be unable to do what it was designed to: represent a "competent and careful" driver in safety-critical cut-ins. This study shows that there is a need for improved modeling and validation of reference driver models—especially those that are to be considered "competent and careful". An analysis of the root cause of the CCDM's failure to perform indicated that the model does not account for the urgency of the situation, although (as previously mentioned) urgency has been shown to play a role in the driver response time in rear-end crashes (Markkula et al., 2016), and it is reasonable to extrapolate that it also plays a similar role in cut-ins. However, for the model to account for urgency, it would have to include a metric related to urgency. One option would be to use lateral POV speed, possibly in addition to lateral TTC. As mentioned in Section 4.2, lateral speed has previously been suggested as a metric for identifying the start of a lane change (Jokhio et al., 2023).

In contrast to the CCDM, the FSM's path prediction seems to facilitate earlier identification of a lane change, but the lateral predicted TTC is still very large at brake initiation and the model can trigger braking much earlier (see Fig. 5E) than human drivers. A path prediction trigger without an urgency component may already trigger braking at the start of a lane change two lanes over. On the other hand, incorporating an urgency-based component—such as lateral TTC—into the algorithm in the FSM and, potentially, in the CCDM (with or without a lateral speed component) could create a response more similar to that of human drivers.

So which aspects of human behavior and performance should the models capture? That of course depends on what it will be used for. The UNECE models are described as models that provide guidance to identify what crashes are preventable and unpreventable by human drivers. Such models should thus at least not result in crashes in situations where a competent and careful human driver avoids crashing. Theoretically, an ADS successfully assessed against a model that fails to fulfill that criterion (as we argue the CCDM does) may be a potential danger in traffic. In the context of UNECE and preventable and unpreventable situations, understanding the safety implications of a model being overly conservative (as our results, as well as the results from Mattas et al., 2022, indicate the FSM is) is more difficult. A benchmark model with overly anticipatory behavior may identify a crash as preventable by a competent and careful human driver in scenarios where humans may not actually be able to prevent a crash. For instance, consider assessing an ADS on an event in which the POV early in the event moves a bit laterally toward the ego vehicle in its own lane—not to change lane but just part of satisficing lane keeping—but then, after a second or two, decided to quickly change into the ego's lane. If a benchmark ego driver model starts braking at the first (mild) lateral movement by the POV (as the FSM does)—a movement to which even a competent and careful human may not react—the benchmark model may avoid a crash even if the competent and careful human driver would not. That is, it may be that the human driver is not able to avoid the quick lateral maneuver, while the benchmark model already moved out of the way of the quickly encroaching POV. If that would be the case, the pass criteria for the ADS safety performance would be unnecessarily strict, demanding safer behavior from the ADS than from a competent and careful human driver. This is yet another indication that competent and careful human benchmark models should not only be validated against everyday driving. Actually, it is an indication that both near-crashes and crashes should be considered in such validation—the latter as it will further help to test the models against the limits of the humans' crash avoidance capabilities.

In summary, we have in this paper shown that it is not enough to validate a competent and careful human benchmark model not only on crash avoidance but also on timing of the brake initiation. That is, an ADS should never crash if a 'competent and careful' model does not, and the ADS should in general react neither later than nor earlier than the benchmark (the exact timing needs to be established in future research).

Given the arguments above, it seems that, if human benchmark models are to be included in regulation, there is a clear need for better and more thoroughly validated models of "competent and careful" drivers. To develop appropriate benchmark models, a deeper understanding of drivers' reaction mechanisms is likely needed—including some way to assess situation urgency. The concept of Comfort Zone Boundaries (CZB; Bärgman et al., 2015; Summala, 2007) may be a feasible way forward. Using CZBs in reference driver models can narrow down the time window for a reaction to an imminent threat, and keep the responses anchored in terms of human's perceptions of safety (in terms of CZBs). In other words, thresholds for a reaction mechanism based on CZBs, obtained through specific experiments, may avoid unnecessarily early interventions, while ensuring a timely reaction to the actual threat. The use of CZBs has been proposed by Olleja et al. (2023) for the specific scenario of a car overtaking a cyclist, but more research is needed to make it practical to implement CZBs in driver models across scenarios—including the ALKS use case. Further, Engström et al. (2024) from Waymo recently presented the *Non-Impaired road user with their Eyes ON the conflict* (NIEON) reference driver model. The NIEON model is a framework that models the response process as an update in beliefs triggered by perceived breaches of prior expectations. Essentially, it operates as a surprise-based system for modeling reference driver behavior. This modelling approach should also be considered in future human benchmark modeling.

Should then regulations (such as the UNECE Regulation 157) include 'competent and careful' human benchmark models? An answer to that question is clearly out of the scope of this work, but having such models be part of regulations is an issue if the models are not really representing what they set out to represent.

## 4.4 Limitations

The limitations of this study are mainly related to the data selection and the annotation processes (covered in Section 4.1). An additional limitation related to the SHRP2 data is occasional poor video quality from the front-facing camera, especially at night. Poor resolution adds to the uncertainty of the POV annotations (and their subsequent transformation to cartesian coordinates). This issue is likely to impact the accuracy of the measurements of POV position and lateral speed for some events (particularly when the POV is farther away), but the effect on the overall study results is likely to be marginal, as the distances between the vehicles are relatively short. As mentioned previously, the radar data were used to improve the video-based data, but they were not always usable, as the radar tracked the objects only intermittently and the data were missing altogether for some events. Therefore, when it was available, it was only used to calibrate the estimated distance between the vehicles based on the manual annotations.

Another limitation, not related to the data, is that the potential influence of other traffic and other factors that may have induced a brake reaction by the ego driver were ignored: for each event, only the ego vehicle and the POV were annotated. The reaction of the ego driver was assumed to be triggered by the safety-critical maneuver of the annotated POV. There may thus be events in which the braking onset of the ego driver was independent of the POV, but we just do not know.

Finally, although ultimately important for a full assessment of traffic safety, an analysis of the risk of injuries caused by the crashes was not performed, as the UNECE regulation specifically addresses only the capability of the models to avoid crashes, and not the potential for injury. If a regulation were to include injury risk in the assessment as well, that perspective should also be validated.

# 5 Conclusions

This study assessed the two computational "competent and careful performance models" of UNECE Regulation 157 (UNECE, 2023) by applying them to 38 safety-critical cut-in near-crashes from the SHRP2 database in virtual simulations. A comparison was made between the models' performances and that of the human drivers in the original SHRP2 events. The metrics were: crash avoidance, timing of brake onset, and distance to the lane at brake onset. It was found that the models performed rather differently than the human drivers. The CCDM showed a delay in the braking response, which in three events actually resulted in a crash that did not occur in the original events. The FSM, on the contrary, showed a more conservative behavior, anticipating the braking response considerably compared to the human drivers. That is, the CCDM model seems to be neither competent nor careful, while the FSM may be overly careful.

In the UNECE regulation these models are limited to providing guidance for the definition of preventable and unpreventable safety-critical driving situations, but we argue that the delayed reaction times (CCDM) and (possibly) overly conservative responses (FSM) may negatively influence the definitions of the boundaries for the safe operation of ALKS. The model characteristics could thus potentially affect both the safety of a "certified" ADS (an ADS that complies with the CCDM might crash in some situations where humans do not) and its sensitivity (an ADS that responds when the FSM responds maybe be too conservative and may be considered a nuisance to the users and a potential hazard for surrounding traffic, as it would not have human-like predictability).

## Acknowledgements

The study was funded by Vinnova (Swedish governmental agency for innovation), the Swedish Energy Agency, the Swedish Transport Administration and involved industry partners, through the strategic vehicle research and innovation program FFI, as part of the project *Improved quantitative driver behavior models and safety assessment methods for ADAS and AD* (QUADRIS: nr. 2020-05156). Gustav Markkula was for this work funded by Volvo Car Corporation and by the UK Engineering and Physical Sciences Research Council (grant EP/S005056/1). The SHRP2 data used in this study has the identifier DOI 10.15787/VTT1/SS1UBM and was made available to us by the Virginia Tech Transportation Institute (VTTI) under the data licence agreement SHRP2-DUL-A-2-18-354. The findings and conclusions of this paper are those of the authors and do not necessarily represent the views of VTTI, the Transportation Research Board (TRB), or the National Academies. We also want to thank Mikael Jung Aust at Volvo Car Corporation for his project management and support.

## Appendix – SHRP2 query

The following describes the selection equivalent in the SHRP2 insight database. However, note that this selection was made on available time-series data in the Data Licensing Agreement (SHRP2-DUL-A-2-18-354)

| Event Severity 1 | Event Nature 1 | Precipitating Event | Pre-Incident Maneuver |
|---|---|---|---|
| • Crash<br>• Near-Crash | • Conflict with vehicle in adjacent lane<br>• Conflict with merging vehicle | • Other vehicle lane change - left in front of subject<br>• Other vehicle lane change - right in front of subject<br>• Other vehicle lane change - left other<br>• Other vehicle lane change - right other<br>Other vehicle from entrance to limited access highway | • Going straight, constant speed<br>• Going straight, accelerating<br>• Decelerating in traffic lane<br>• Changing lanes |

*Table 1 – Filtering criteria for the SHRP2 events selection, available at https://insight.shrp2nds.us. Query ID 155692.*